\documentclass{moriond}

\usepackage{ amssymb }
\usepackage{ amsmath }
\usepackage{ relsize }
\usepackage{ hyperref }

\def\be{\begin{equation}}
\def\ee{\end{equation}}
\def\bea{\begin{eqnarray}}
\def\eea{\end{eqnarray}}

\newcommand{\Photo}{}
\newcommand{\tightoverset}[2]{%
  \mathop{#2}\limits^{\vbox to -.5ex{\kern-0.75ex\hbox{$#1$}\vss}}}
\newcommand{\pt}{\mbox{$p_{\mathrm{T}}$}}
\newcommand{\vn}{\mbox{$v_n$}}
\newcommand{\vnRP}{\mbox{$v_n^\mathrm{RP}$}}
\newcommand{\vtwoRP}{\mbox{$v_2^\mathrm{RP}$}}

\newcommand{\vvn}{{\tightoverset{\mathsmaller{\mathsmaller{\rightarrow}}}{v}}_{\! n}}

\newcommand{\vvnRP}{{\tightoverset{\mathsmaller{\mathsmaller{\rightarrow}}}{v}}_{\! n}^{ \mathrm{RP}}}

\newcommand{\Phin}{\mbox{$\Phi_n$}}
\newcommand{\fd}{\mathrm{d}}
\newcommand{\sqrtsnn}{\mbox{$\sqrt{s_{\mathrm{NN}}}$}}
\newcommand{\mub}{\mbox{$\mu \mathrm{b}^{-1}$}}
\newcommand{\pp}{\mbox{$p$+$p$}}
\newcommand{\pPb}{\mbox{\it p}+Pb}
\newcommand{\PbPb}{\mbox{Pb+Pb}}
\newcommand{\SEt}{\mbox{$E_{\mathrm{T}}^{\mathrm{Pb}}$}}
\def\<{\left\langle}
\def\>{\right\rangle}

\begin{document}
\vspace*{4cm}
\title{Elliptic flow phenomenon at ATLAS}

\author{ Martin Spousta on behalf of the ATLAS Collaboration  }

\address{ $^{\ast}$Charles University in Prague, Institute of Particle and Nuclear Physics, \\ V Holesovickach 2, 180 00 Prague 8, Czech republic}

\maketitle


\abstract{
  We summarize measurements of elliptic flow and higher order flow harmonics performed by the ATLAS experiment at 
the LHC. Results on event-averaged flow measurements and event-plane correlations in \PbPb\ collisions are 
discussed along with the event-by-event flow measurements. Further, we summarize results on flow in 
\pPb\ collisions.
}

\section{Introduction}
\label{sec:1}

Heavy ion collisions at the Relativistic Heavy Ion Collider (RHIC) and the Large Hadron Collider (LHC) 
create hot and dense matter that is composed of deconfined quarks and gluons. A useful tool to study 
properties of this matter is the azimuthal anisotropy of particle emission. At low transverse momenta ($\pt \lesssim 3-4$~GeV), 
this anisotropy results from a pressure driven anisotropic expansion of the created matter, with more 
particles emitted in the direction of the largest pressure gradient~\cite{Ollitrault:1992bk}. At higher 
\pt\, this anisotropy is understood to result from the path-length dependent energy loss of jets as they 
traverse the matter, with more particles emitted in the direction of smallest path-length 
\cite{Gyulassy:2000gk}. These directions of maximum emission are strongly correlated, and the observed 
azimuthal anisotropy can be expressed~\cite{Poskanzer:1998yz} as a Fourier series in azimuthal angle $\phi$,
\be
\frac{1}{2\pi \pt} \frac{\fd^3 N}{\fd \phi \fd \pt \fd \eta} =
\frac{1}{2\pi \pt} \frac{\fd^2 N}{\fd \pt \fd \eta}\left( 1 + \sum_{n=1}^{\infty} \vn(\pt, \eta) \cos n (\phi - \Phin) \right),
\label{eqn1}
\ee
  where 
  $\eta$ is pseudorapidity, \vn\ and \Phin\ represent the magnitude 
and direction of the $n^\mathrm{th}$-order harmonic, respectively. The $n^\mathrm{th}$-order harmonic has 
$n$-fold periodicity in azimuth, and the coefficients at low \pt\ are often given descriptive names, such as 
``direct flow'' ($v_1$), ``elliptic flow'' ($v_2$), or ``triangular flow'' ($v_3$). 

  In typical non-central\,\footnote{In \PbPb\ collisions, the centrality is estimated using energy deposited 
in forward ($3.1<|\eta|<4.9$) calorimeters.} heavy ion collisions, the large and dominating $v_2$ is associated 
mainly with the almond shape of the nuclear overlap. However, $v_2$ in central (head-on) collisions 
and the other $v_n$ coefficients in general are related to various shape components of the initial state 
arising from fluctuations of the nucleon positions in the overlap region~\cite{Alver:2010gr}. 

  The event-averaged measurement of elliptic flow and higher order harmonics in \PbPb\ collisions is 
summarized in Sec.~\ref{sec:2}. The event-by-event flow measurements in \PbPb\ collisions are discussed in 
Sec.~\ref{sec:3}. Section~\ref{sec:4} summarizes results from two-particle correlations and flow in \pPb\ 
collisions. Measurements in \PbPb\ collisions use the data at $\sqrtsnn = 2.76$~TeV with
integrated luminosity of 7 \mub. Measurements in \pPb\ collisions use the data at $\sqrtsnn = 5.02$~TeV with 
integrated luminosity of approximately 1 \mub.

\section{Elliptic flow and higher order harmonics in \PbPb\ collisions}
\label{sec:2}

The elliptic flow and higher order harmonics in \PbPb\ collisions were extracted from ``event-plane 
method'' \cite{ATLAS:2011ah,ATLAS:2012at} and using multi-particle azimuthal correlations 
\cite{ATLAS-CONF-2012-118,ATLAS:2012at}. The event-plane method correlates individual tracks with the 
event-plane direction \Phin\ measured using energy deposited in the forward calorimeters. The Fourier 
coefficient \vn\ can be expressed as
\be
  \vn \equiv \< \cos[ n( \phi - \Phin ) ] \>,
\label{eqn2}
\ee
  where angle brackets denote two-step averaging, first over charged particles in an event, and then over 
events. 

  Significant $v_2$--$v_6$ values were observed as a function of transverse
momentum ($0.5 < \pt < 20$~GeV), pseudorapidity ($|\eta| < 2.5$), and centrality.
  All flow harmonics exhibit similar dependence on \pt. 
  The values of \vn\ first grow with increasing \pt\ achieving a maximum at around 3~GeV, then they 
decrease staying non-zero across the whole measured \pt\ interval.
  The \pt\ dependence of \vn\ values for $n \geq 3$ is found to follow an approximate scaling relation, 
$v_n^{1/n}(\pt) \sim v_2^{1/2} (\pt)$, except in the top 5\% most central collisions.
  The values of \vn\ for $n \geq 2$ do not exhibit a significant variation when 
evaluated as a function of $\eta$.
  While a similar \pt\ dependence of flow was previously observed at RHIC, the $\eta$ dependence of flow at 
RHIC was different, achieving a maximum at mid-rapidity and decreasing with increasing $\eta$ 
\cite{Back:2004mh}.

  The centrality dependence of  $v_2$ reflects the geometry of the collision. It achieves the 
maximum in mid-central collisions where the ellipticity of the initial overlapping region is largest. The 
$v_2$ decreases in more central collisions due to decreasing initial eccentricity and it also decreases in 
more peripheral collisions due to lack of collectivity~\cite{Kolb:2001qz}.
  Contrary to the behavior of $v_2$, the \vn\ values for $n \geq 3$ are found to vary only weakly with centrality.

  The basic conclusions derived from results obtained using the event-plane method are consistent with the 
conclusions from results obtained using multi-particle azimuthal correlations.
  In the event-plane method, only two-particle correlations are exploited in the determination of \vn\ (see 
Eq.\ref{eqn2}). This leads to a well-known problem of disentangling all-particle flow and contributions 
from particle correlations unrelated to the initial geometry, known as non-flow correlations. These 
non-flow effects include correlations due to energy and momentum conservation, resonance decays, quantum 
interference phenomena, and jet production. In order to suppress non-flow correlations, methods that use 
genuine multi-particle correlations can be employed. 
  Two particle correlations allows to determine Fourier coefficient \vn\ without estimating the 
event-plane direction as follows
\be
\< {corr}_n \{2\} \> \equiv \< \exp [in(\phi_1 - \phi_2 )] \> = \< \cos [n(\phi_1 - \phi_2) ] \> = v_n\{2\}^2,
\ee
  where $\phi_1$ and $\phi_2$ denote azimuthal angles of two particles forming a pair. Angle brackets denote 
two-step averaging, same as in the case of determining the \vn\ coefficients using the event-plane method.
  This can be generalized to $2k$-particle correlations defined as
\be
\< {corr}_n \{2k\} \> 
                             =      \< \exp [ in( \phi_1 + ... - \phi_{1+k} - ... - \phi_{2k} )] \> = v_n\{2k\}^{2k}.
\ee
  The multi-particle correlations $\< {corr}_n \{2k\} \>$ account for the collective anisotropic 
flow as well as for the non-flow effects. The anisotropic flow related to the initial geometry is a global, 
collective effect involving correlations between all outgoing particles. Thus, in absence of non-flow 
effects, $\vn\{2k\}$ is expected to be independent of $k$. On the contrary, most of the non-flow effects
are contributing to correlations of few particles only. Thus, $2k$-particle ``cumulants'' can be 
used to suppress the non-flow contribution by eliminating the correlations between fewer than $2k$ 
particles. An example is the cumulant of the four-particle correlations, 
  $c_n\{4\} \equiv {corr}_n\{4\} - 2 {corr}_n\{2\}^2$, which measures the genuine 
four-particle correlations. 
  If the non-flow contribution is only due to the two-particle correlations, then 
they are eliminated and $c_n\{4\}$ directly measures flow harmonics. In practice, multi-particle azimuthal 
correlations are calculated using the generating functions formalism~\cite{Borghini:2001zr}.

  The cumulant approach to measure flow harmonics also provides a possibility to study elliptic flow 
fluctuations which can be related to the fluctuations in the initial geometry of the interaction region 
\cite{Alver:2007qw}. The prediction for the event-by-event variation in the initial geometry obtained 
from the Glauber Monte Carlo model~\cite{Broniowski:2007nz}, shows a similar size of fluctuations, 
suggesting that the elliptic flow fluctuations could originate from fluctuations in the initial geometry.


\section{Event-by-event measurement of flow}
\label{sec:3}

The flow signal is clearly visible on the event-by-event basis in \PbPb\ collisions. Yield of particles in 
one event evaluated as a function of azimuthal angle can vary by as much as a factor of 
three~\cite{Aad:2013xma}. Measurement of event-by-event flow coefficients allows to directly access the flow 
fluctuations and thus to better understand the role of initial geometry in forming the flow effects. The 
event-by-event flow can be quantified by the per-particle ``flow-vector'', $\vvn = (\vn \cos n \Phin, \vn 
\sin n \Phin$).
  If fluctuations of $\vvn$ relative to the flow vector associated with the average geometry in the reaction plane~\footnote{Reaction plane is defined by the impact parameter vector and the beam axis. Reaction plane needs to be distinguished from the event plane \Phin\ which is directly accessible event-by-event.} \!(RP), 
$\vvnRP$, are described by a two dimensional (2D) Gaussian function in the 
transverse plane,\cite{Voloshin:2007pc} then the probability density of $\vvn$ can be expressed as
\vspace{-1em}
\be
p(\vvn) = \frac{1}{2 \pi \delta_{v_n}^2} e^{ (\vvn - \vvnRP)^2/(2\delta_{v_n}^2)}.
\ee
  %
  The relation between the event-averaged flow coefficients $\< \vn \>$ discussed in Sec.~\ref{sec:2} and 
event-by-event flow coefficients \vn\ can be then written~\cite{Aad:2013xma} as $(\vnRP)^2 \approx \< \vn \>^2 - 
\delta_{v_n}^2$.

 The flow coefficients \vn\ were 
measured for $n=2,3,$ and 4 over the pseudorapidity range $|\eta|<2.5$ and the transverse momentum range 
$\pt > 0.5$~GeV~\cite{Aad:2013xma}. 
  In the very central, \mbox{0--2\%}, collisions, where the eccentricity of the initial overlapping region 
approaches zero, the measured $v_2$ distributions are found to approach that of a radial projection of a 2D 
Gaussian distribution centered around zero ($\vtwoRP = 0$). This is consistent with a scenario where 
fluctuations are the primary contribution to the overall shape for these most central collisions.
  Starting with the centrality interval 5--10\%, the $v_2$ distributions differ significantly from this 
scenario, suggesting that they have a significant component associated with the average collision geometry. 
In contrast, the $v_3$ and $v_4$ are consistent with a pure 2D Gaussian-fluctuation scenario (i.e. $\vnRP = 
0$) over most of the measured centrality range. However, a systematic deviation from this fluctuation-only 
scenario is observed for $v_3$ and $v_4$ in mid-central collisions.

  The \vn\ distributions were also measured separately for charge particles with $0.5 < \pt < 1$~GeV and 
$\pt > 1$~GeV. The shape of the unfolded distributions, when rescaled to the same $\< \vn \>$, is found to 
be nearly the same for the two \pt\ ranges. This suggests that the hydrodynamic response to the eccentricity 
of the initial geometry has little variation in this \pt\ region. The conclusions were quantified in more 
details e.g. by evaluating the ratios of width to the mean of measured \vn\ distributions. Further, the measured 
\vn\ distributions were compared with the eccentricity distributions of the initial geometry from the 
Glauber model~\cite{Miller:2007ri} and MC-KLN model~\cite{Drescher:2006pi}. Both models were found to fail 
in describing the data consistently over most of the measured centrality range.

  More insight to the role of fluctuations in flow effects can be also gained by measuring the event-plane 
correlations.
  If the fluctuations in initial geometry are small and random, the orientations of event-plane directions 
\Phin\ of different order are expected to be uncorrelated.
  Fourteen correlators of event-plane directions were measured~\cite{Aad:2014fla} using a standard 
event-plane method and a scalar-product method~\cite{Adler:2002pu}.
  Several different trends in the centrality dependence of these correlators were observed. These trends 
were not reproduced by predictions based on the Glauber model, which includes only the correlations from 
the collision geometry in the initial state. Calculations that include the final state collective dynamics 
are able to describe qualitatively, and in some cases also quantitatively, the centrality dependence of 
the measured correlators.
  In particular, the AMPT model~\cite{Lin:2004en} which generates collective flow by elastic scatterings in 
the partonic and hadronic phase was shown to reproduce the trends seen in the data.
  These observations suggest that both the fluctuations in the initial geometry and non-linear mixing 
between different harmonics in the final state are important for creating the correlations in the momentum 
space.


\section{Flow in \pPb\ collisions}
\label{sec:4}

High-multiplicity \pPb\ events provide a rich environment for studying observables associated with high 
parton densities in hadronic collisions. Tool to probe this physics is the two-particle correlation 
function measured in terms of the relative pseudorapidity ($\Delta \eta$) and azimuthal angle ($\Delta 
\phi$) of selected particle pairs, $C(\Delta \eta, \Delta \phi)$. The first studies of two-particle 
correlations in the highest-multiplicity \pp\ collisions at the LHC~\cite{Khachatryan:2010gv} showed an 
enhanced production of pairs of particles at $\Delta \phi \sim 0$, with the correlation extending over a 
wide range in $\Delta \eta$, a feature frequently referred to as a ``ridge.''

  Similar long range ($2 < |\Delta \eta| < 5$) correlations were observed in \pPb\ collisions on the 
near-side ($\Delta \phi \sim 0$) exhibiting a rapid grow with increasing event activity characterized by 
the transverse energy (\SEt ) summed over $3.1 < \eta < 4.9$ in the direction of the Pb beam. Further, a 
long-range away-side ($\Delta \phi \sim \pi$) correlation was found to be present in high-event activity 
\pPb\ collisions after subtracting the expected contributions from recoiling dijets and other sources 
estimated using events with small \SEt ~\cite{Aad:2012gla}. In this measurement, the correlation function 
$C(\Delta \eta, \Delta \phi)$ was also projected to the $\Delta \phi$ direction. The resultant $\Delta 
\phi$ correlation was found to be approximately symmetric about $\pi/2$ exhibiting thus a clear flow 
signal. To quantify the size of the flow signal, $v_2$ is determined using two- and four-particle 
cumulants~\cite{Aad:2013fja}. A significant magnitude of $v_2$ is observed for both $v_2\{2\}$ and 
$v_2\{4\}$, although $v_2\{2\}$ is consistently larger than $v_2\{4\}$, indicating a sizable contribution 
of non-flow correlations to $v_2\{2\}$. The transverse momentum dependence of $v_2\{4\}$ shows a behavior 
similar to that measured in \PbPb\ collisions.

  Presence of flow in \pPb\ collisions might have not been expected due to the small size of the produced 
system compared to the mean free path of interacting constituents. Despite that, it was observed that the 
prediction of viscous hydrodynamics can reproduce the magnitude of the measured flow when configured 
with similar initial conditions as those used for \PbPb\ collisions~\cite{Aad:2013fja,Bozek:2012gr}.
  Many of the physics mechanisms proposed to explain the \pp\ ridge, including multi-parton interactions, 
parton saturation, and collective expansion of the final state, are also expected to be relevant in \pPb\ 
collisions and they may contribute to the observed flow. The flow phenomenon in \pPb\ collisions thus 
clearly deserves more investigations to understand its origin.





\section*{Acknowledgments}

This work was supported by Charles University in Prague, projects INGOII LG13009, PRVOUK P45, and UNCE 204020/2012.

\end{document}